\documentclass[aps,prc,twocolumn,showpacs,superscriptaddress,amsmath,amssymb,nofootinbib]{revtex4}
\usepackage{graphicx}% Include figure files
\usepackage{dcolumn}% Align table columns on decimal point
\usepackage{bm}% bold math
\newcommand{\nuclei}[2]{\ensuremath{\mathrm{^{#1}#2}}}
\newcommand{\helium}[1][4]{\nuclei{#1}{He}}
\newcommand{\carbon}[1][12]{\nuclei{#1}{C}}
\newcommand{\nitrogen}[1][14]{\nuclei{#1}{N}}
\newcommand{\oxygen}[1][16]{\nuclei{#1}{O}}
\newcommand{\neon}[1][20]{\nuclei{#1}{Ne}}
\newcommand{\sodium}[1][23]{\nuclei{#1}{Na}}

\newcommand{\nickel}[1][58]{\nuclei{#1}{Ni}}

\begin{document}

\preprint{APS/123-QED}

\title{Gamow-Teller strength for the analog transitions to the first T=1/2, $J^{\pi}=3/2^{-}$ states in $^{13}$C and $^{13}$N and the implications for Type Ia supernovae.}

\author{R.G.T. Zegers}
\email{zegers@nscl.msu.edu} \affiliation{National Superconducting Cyclotron Laboratory, Michigan
State University, East Lansing, MI 48824-1321, USA} \affiliation{Department of Physics and
Astronomy, Michigan State University, East Lansing, MI 48824, USA} \affiliation{Joint Institute for
Nuclear Astrophysics, Michigan State University, East Lansing, MI 48824, USA}
\author{E. F. Brown}
\affiliation{National Superconducting Cyclotron Laboratory, Michigan State University, East
Lansing, MI 48824-1321, USA} \affiliation{Department of Physics and Astronomy, Michigan State
University, East Lansing, MI 48824, USA} \affiliation{Joint Institute for Nuclear Astrophysics,
Michigan State University, East Lansing, MI 48824, USA}
\author{H. Akimune}
\affiliation{Department of Physics, Konan University, Kobe, Hyogo, 658-8501, Japan}
\author{Sam M. Austin}
\affiliation{National Superconducting Cyclotron Laboratory, Michigan State University, East Lansing, MI 48824-1321, USA}
\affiliation{Joint Institute for Nuclear Astrophysics, Michigan State University, East Lansing, MI 48824, USA}
\author{A.M. van den Berg}
\affiliation{Kernfysisch Versneller Instituut, University of Groningen, Zernikelaan 25, 9747 AA
Groningen, The Netherlands}
\author{B.A. Brown}
\affiliation{National Superconducting Cyclotron Laboratory, Michigan State University, East
Lansing, MI 48824-1321, USA} \affiliation{Department of Physics and Astronomy, Michigan State
University, East Lansing, MI 48824, USA} \affiliation{Joint Institute for Nuclear Astrophysics,
Michigan State University, East Lansing, MI 48824, USA}
\author{D. A. Chamulak}
\affiliation{Department of Physics and Astronomy, Michigan State University, East
Lansing, MI 48824, USA} \affiliation{Joint Institute for Nuclear Astrophysics,
Michigan State University, East Lansing, MI 48824, USA}
\author{Y. Fujita}
\affiliation{Department of Physics, Osaka University, Toyonaka, Osaka 560-0043, Japan}
\author{M. Fujiwara}
\affiliation{Kansai Photon Science Institute, Japan Atomic Energy Agency, Kizu, Kyoto 619-0215, Japan}
\affiliation{Research Center for Nuclear Physics, Osaka University, Ibaraki, Osaka 567-0047, Japan}
\author{S. Gal\`{e}s}
\affiliation{Institut de Physique Nucl\'{e}aire, IN2P3-CNRS, Orsay, France}
\author{M.N. Harakeh}
\affiliation{Kernfysisch Versneller Instituut, University of Groningen, Zernikelaan 25, 9747 AA Groningen, The Netherlands}%
\author{H. Hashimoto}
\affiliation{Research Center for Nuclear Physics, Osaka University, Ibaraki, Osaka 567-0047, Japan}
\author{R. Hayami}
\affiliation{Department of Physics, University of Tokushima, Tokushima 770-8502, Japan}
\author{G.W. Hitt}
\affiliation{National Superconducting Cyclotron Laboratory, Michigan State University, East
Lansing, MI 48824-1321, USA} \affiliation{Department of Physics and Astronomy, Michigan State
University, East Lansing, MI 48824, USA} \affiliation{Joint Institute for Nuclear Astrophysics,
Michigan State University, East Lansing, MI 48824, USA}
\author{M. Itoh}
\affiliation{Cyclotron and Radioisotope Center, Tohoku University, Sendai, Miyagi 980-8578, Japan}
\author{T. Kawabata}
\affiliation{Center for Nuclear Study, University of Tokyo, RIKEN Campus, Wako, Saitama 351-0198, Japan}
\author{K. Kawase}
\affiliation{Research Center for Nuclear Physics, Osaka University, Ibaraki, Osaka 567-0047, Japan}
\author{M. Kinoshita}
\affiliation{Department of Physics, Konan University, Kobe, Hyogo, 658-8501, Japan}
\author{K. Nakanishi}
\affiliation{Research Center for Nuclear Physics, Osaka University, Ibaraki, Osaka 567-0047, Japan}
\author{S. Nakayama}
\affiliation{Department of Physics, University of Tokushima, Tokushima 770-8502, Japan}
\author{S. Okumura}
\affiliation{Research Center for Nuclear Physics, Osaka University, Ibaraki, Osaka
567-0047, Japan}
\author{Y. Shimbara}
\affiliation{National Superconducting Cyclotron Laboratory, Michigan State University, East Lansing, MI 48824-1321, USA}
\affiliation{Joint Institute for Nuclear Astrophysics, Michigan State University, East Lansing, MI 48824, USA}
\author{M. Uchida}
\affiliation{Tokyo Institute of Technology, 2-12-1 O-Okayama, Tokyo 152-8550, Japan}
\author{H. Ueno}
\affiliation{Applied Nuclear Physics Laboratory, RIKEN, Wako, Saitama 351-0198, Japan}
\author{T. Yamagata}
\affiliation{Department of Physics, Konan University, Kobe, Hyogo, 658-8501, Japan}
\author{M. Yosoi}
\affiliation{Research Center for Nuclear Physics, Osaka University, Ibaraki, Osaka 567-0047, Japan}
\date{\today}%

\begin{abstract}
The Gamow-Teller strength for the transition from the ground state of $^{13}$C to
the T=1/2, $J^{\pi}=3/2^{-}$ excited state at 3.51 MeV in $^{13}$N is extracted via
the $^{13}$C($^{3}$He,$t$) reaction at 420 MeV. In contrast to results from earlier
($p$,$n$) studies on $^{13}$C, a good agreement with shell-model calculations and
the empirical unit cross section systematics from other nuclei is found. The results
are used to study the analog $\nitrogen[13](e^{-},\nu_{e})\carbon[13]$ reaction,
which plays a role in the pre-explosion convective phase of type Ia supernovae.
Although the differences between the results from the ($^{3}$He,$t$) and ($p$,$n$)
data significantly affect the deduced electron-capture rate and the net
heat-deposition in the star due to this transition, the overall effect on the
pre-explosive evolution is small.
\end{abstract}

\maketitle

\section{Introduction}
\label{sec:intro}

Type Ia supernovae (SNe Ia) are thought to be thermonuclear incinerations of
carbon-oxygen white dwarf stars after accretion from a companion has increased their
mass to near the Chandrasekar limit (for a review, see e.g. \cite{HIL00}). It is
unclear, however, what type of binary systems produce SNe Ia and how the structure
of the progenitor affects the outcome of the explosion. Observational studies show that the peak brightness of SNe Ia, and hence the mass of \nickel[56] synthesized in the explosion, are correlated with the properties of the host galaxy \cite{Sullivan2006Rates-and-prope}.  One proposed explanation for this is that the ratio of \nickel[56] to stable nickel and iron isotopes depends on the electron abundance $Y_{e}$ in the white dwarf at the time of the explosion. A reduction in $Y_{e}$
shifts the isotopic distribution of iron-peak ejecta to more neutron-rich isotopes \cite{BRA00}
and decreases the mass of \nickel[56] ejected, which in turn reduces the peak
brightness of the supernova. In the pre-explosion white dwarf, $Y_{e}$ is set by the abundance of trace nuclides such as \neon[22] \cite{timmes.brown.ea:variations}.

In recent works \cite{CHA07,PIR07} it was pointed out that reactions during the pre-explosion
``simmering'' phase (during the $\sim 10^{3}\,\text{yr}$ prior to the explosion)
also reduce $Y_{e}$.  During this period, protons formed from
$\carbon(\carbon,p)\sodium$ predominantly capture onto \carbon\ to form
$\beta$-unstable \nitrogen[13]. The $\nitrogen[13](e^{-},\nu_{e})\carbon[13]$
reaction thus plays a role. Since the stellar density is $\rho=(1\text{--}3) \times 10^{9}$, the
electron Fermi energy is sufficiently high ($\approx 5\,\text{MeV}$) to allow
electron capture on the \nitrogen[13] ground state (spin-parity $J^{\pi}=1/2^{-}$)
via an allowed Gamow-Teller (GT) transition into the $J^{\pi}=3/2^{-}$ state of
\carbon[13] at $3.68\,\text{MeV}$ \cite{CHA07}, in addition to the capture to the
\carbon[13] ground state ($J^{\pi}=1/2^{-}$). Note that captures into states at higher excitation energies are not important for the range of densities found in white dwarf stars, so the heating from this reaction is set by captures into this excited state and the ground state.
As discussed briefly in Ref. \cite{CHA07}, where a shell model calculation was used to estimate the contribution of the excited state, the increased capture rate does not strongly effect the reduction of $Y_{e}$, which is instead governed by the production of \nitrogen[13], not by the faster (under these conditions) electron capture. The capture into the excited state does contribute to the effective heat per $\carbon+\carbon$ reaction during simmering, however, and so a more accurate determination of the branching into the excited state is useful for future studies of the simmering stage of SNe Ia, including those focusing on the hydrodynamics. Hydrodynamical simulations of the simmering phase \cite{Woosley2004Carbon-Ignition,Kuhlen2006Carbon-Ignition} require an
accurate determination of the effective heat released per $\mathrm{^{12}C + ^{12}C}$ reaction, since this heating sets the amount of $\mathrm{^{12}C}$ consumed, and hence the neutronization of the white dwarf, prior to the explosion. Motivated by the previous work \cite{CHA07}, this paper incorporates experimental data into the determination of the electron capture rate into the excited state and the related increase in heat deposition.

The weak transition strength from the $^{13}$N ground state to the $^{13}$C ground state is well
determined using the experimental $\log ft$ value of $3.6648\pm0.0005$ for $\beta^{+}$-decay
\cite{AJZ91}. Because the excitation energy of the 3.68 MeV state is higher than the Q-value for
$\beta^{+}$-decay (2.22 MeV), no such data are available for the transition to this state.
Therefore, in Ref. \cite{CHA07}, the GT transition strength from the $^{13}$N ground state to the
excited state at 3.68 MeV in $^{13}$C was calculated (B(GT)=1.5) using the shell-model code OXBASH
\cite{OXBA} employing the CKII interaction \cite{COH67} in the $p$-shell model space, taking into
account a phenomenological quenching factor of 0.67 \cite{CHO93} for the Gamow-Teller strength
\footnote{This value is calculated by squaring the quenching factor of the free-nucleon operator
given by $q_{s}=1-0.19(\frac{A}{16})^{0.35}$ \cite{CHO93}.}.

\begin{figure}
\includegraphics[width=8.0cm]{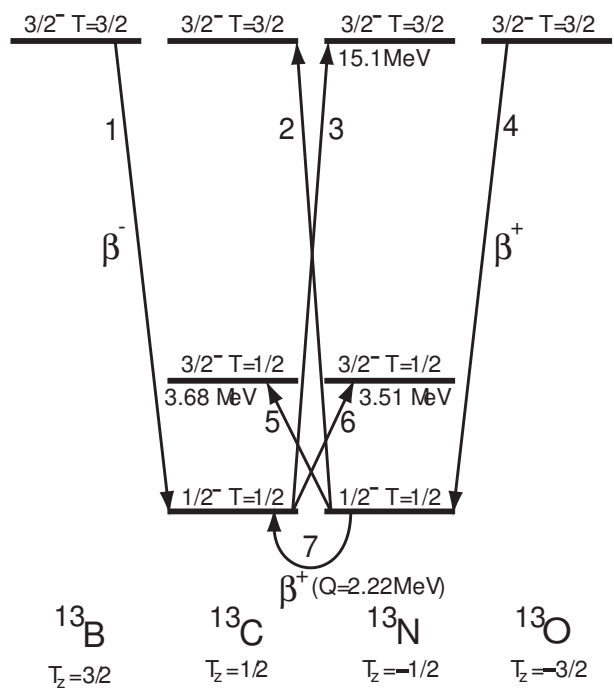}
\caption{\label{levels} Isospin analogous transitions in the A=13 and
$T_{z}=\pm1/2,\pm3/2$ isobar system are schematically shown. The Coulomb
displacement energies are set to zero so that the isospin symmetry of the system
becomes clearer. For each state, spin-parity $J^{\pi}$ and isospin $T$ are
indicated. Excitation energies are given, where relevant for the present work,
relative to the ground state of the nuclei. Transitions labeled by 1-4 are analogs
and transitions with labels 5 and 6 are analogs. For transitions labeled with
$\beta^{\pm}$, the log$ft$'s are known from $\beta$-decay experiments.}
\end{figure}

Experimental information about the Gamow-Teller strength can be extracted, however,
by studying the analog transition from the $^{13}$C ground state to the
$J^{\pi}=3/2^{-}$ state at 3.51 MeV in $^{13}$N, using a ($p$,$n$)-type
charge-exchange reaction and assuming isospin symmetry (the validity of this
assumption is discussed below). In the simplified level-scheme of relevant A=13
nuclei depicted in Fig. \ref{levels}, these analog transitions are labeled 5 and 6.

The charge-exchange reaction studies rely on the proportionality between B(GT) and the
charge-exchange cross section at zero momentum transfer ($q=0$) (see details below). Several
experiments on $^{13}$C have been performed using the ($p$,$n$) reaction at beam energies between
120 MeV and 200 MeV. In the earlier experiments \cite{TAD82,GOO85,TAD87}, a B(GT) of $0.82\pm0.05$
\footnote{This is the value from Ref. \cite{TAD87}. The B(GT) values in Refs. \cite{TAD82,GOO85}
are within the uncertainties consistent with this value} was extracted. The unusually large
discrepancy between the experimental value and the theoretical one (B(GT)=1.5, taking into account
quenching) was the topic of considerable debate \cite{TAD87,GOO85,WAT85}. More recently, a
measurement of the $^{13}$C($p$,$n$) reaction at 197 MeV gave B(GT)=1.06$\pm$0.05 \cite{WAN01}.
This value is closer but still significantly lower than the shell-model predictions. Because of the significance of this transition for the calculations performed for SNe Ia, further investigation is warranted.

As an alternative to the $^{13}$C($p$,$n$) reaction, we studied the $^{13}$C($^{3}$He,$t$) reaction
at 420 MeV and combined the results with an earlier $^{13}$C($^{3}$He,$t$) experiment performed at
450 MeV \cite{FUJ04}. The data at 420 MeV were taken as part of a program to study the extraction
of GT strength using the ($^{3}$He,$t$) probe \cite{ZEG07} over a wide mass range. Since the
targets used contained natural carbon (1.07\% $^{13}$C) only transitions to low-lying states in
$^{13}$N could be studied. The transition to the $T=3/2, J^{\pi}=3/2^{-}$ state at 15.1 MeV in
$^{13}$N is important for strength calibration purposes. However, in targets made of natural carbon
this state is inseparable from the strong $^{12}$C($^{3}$He,$t$)$^{12}$N(g.s.) reaction since the
Q-values are nearly identical. Therefore, the data from the older experiment, in which a
highly-enriched $^{13}$C target was used, was needed as well. Comparing the two sets also provides
a good way to evaluate systematic errors in the extraction of cross sections.

This paper is structured as follows. After briefly describing the method of
extracting GT strengths from charge-exchange data and the particular methods used
for the A=13 analog transitions, the findings from the $^{13}$C($^{3}$He,$t$)
experiments are presented and compared with the shell-model calculations. It is
shown that the results from the $^{13}$C($^{3}$He,$t$) experiments and the
shell-model calculations are consistent. We then briefly discuss the comparison with
the ($p$,$n$) results. Finally, we estimate the uncertainties in the
electron-capture rates on $^{13}$N in the stellar environment and discuss the
ramifications for the simmering stage of SNe Ia.

\section{Extraction of weak transition strength from the $^{13}$C($^{3}$H\MakeTextLowercase{e},$t$) reaction.}
\label{sec:extra} For both the ($p$,$n$) and ($^{3}$He,$t$) reactions, the
extraction of B(GT) values from the data is based on their proportionality to the
cross section at zero momentum transfer ($q=0$) at sufficiently high beam energies
($\gtrsim 100$ MeV/nucleon). In eikonal approximation, this proportionality is
written as \cite{TAD87}:
\begin{equation}
\label{eq:eik}
\frac{d\sigma}{d\Omega}(q=0)=KN_{\sigma\tau}|J_{\sigma\tau}|^{2}B(GT)=\hat{\sigma}_{\sigma\tau}B(GT).
\end{equation}
Here, $K=\frac{E_{i}E_{f}}{(\hbar^{2}c^{2}\pi)^{2}}$ where $E_{i(f)}$ is the reduced
energy in the incoming (outgoing) channel, $N_{\sigma\tau}$ is the distortion factor
defined by the ratio of the distorted-wave to the plane-wave cross sections;
$|J_{\sigma\tau}|$ is the volume-integral of the central $\sigma\tau$ interaction.
The factor $KN|J_{\sigma\tau}|^{2}$ is referred to as the unit cross section,
$\hat{\sigma}_{\sigma\tau}$. For excitations from a particular target nucleus, the
unit cross section is usually determined experimentally, using transitions for which
the B(GT) is empirically known from $\beta$-decay.

The proportionality of Eq. \ref{eq:eik} is not perfect. In particular, contributions from
incoherent and coherent $\Delta$L=2, $\Delta$S=1 to GT transitions lead to deviations. The
incoherent contribution, associated with a total angular momentum transfer to the relative motion
between the target and projectile $\Delta$J$_{R}$= 2, can be extracted from the data, since its
angular distribution peaks at finite angle and has a minimum at 0$^{\circ}$. Usually, the
$\Delta$J$_{R}=2$ contribution to the 0$^{\circ}$ cross section is no more than a few percent of
the total cross section. The coherent contribution, largely due to the non-central
tensor-interaction and the largest source that breaks the proportionality of Eq. \ref{eq:eik}
\cite{ZEG06,COL06,FUJ07}, cannot easily be determined from the data since it does not strongly
affect the angular distribution at forward scattering angles. Its effect on the cross section must
be estimated by theory. Such estimates have shown to provide reasonable estimates for the breaking
of proportionality for ($^{3}$He,$t$) reactions on $^{26}$Mg \cite{ZEG06} and $^{58}$Ni
\cite{COL06}.

For Fermi transitions, a relationship similar to Eq. \ref{eq:eik} exists between
$\frac{d\sigma}{d\Omega}(q=0)$ and the Fermi strength B(F) with unit cross section
$\hat{\sigma}_{\tau}$, but with a distortion factor $N_{\tau}$ and volume integral
of the central $\tau$ interaction $|J_{\tau}|$. Since no spin transfer is involved,
breaking of the proportionality due to $\Delta$L=2 contributions is not an issue.

Here, we are particularly interested in the excitation of the $3/2^{-}$ state at 3.51 MeV
(transition 6 in Fig. \ref{levels}). The unit cross section used to convert the cross section at
$q=0$ to strength can be calibrated in two ways. The first method relies on the use of the
transition to the $^{13}$N ground state, for which the strength is known from $\beta^{+}$ decay
data \cite{AJZ91}. This transition contains both a Fermi component (non-spin-flip, $\Delta$S=0) and
GT component ($\Delta$S=1). Under the assumption that the Fermi strength $B(F)$ exhausts the full
Fermi sum rule ($S_{-}(F)-S_{+}(F)=(N-Z)=1$) \footnote{This assumption is valid on the level of
0.15(9)\% \cite{ORM95}}, the strength of the GT component $B(GT)=0.207\pm0.002$ can be found using:
\begin{equation}
\label{eq:logft} B(F)+\left(\frac{g_{A}}{g_{V}}\right)^{2}B(GT)=\frac{K/g^{2}_{V}}{ft},
\end{equation}
with $(\frac{g_{A}}{g_{V}})=1.264\pm0.002$ \cite{DUB91} and $K/g^{2}_{V}=6147\pm7$
\cite{HAR90}. However, to extract the unit cross section for the GT component only,
the unit cross section for the Fermi component must be known with good accuracy,
which makes this transition harder to use as a calibration tool.

Fortunately, a second method to independently calibrate the GT unit cross section is available,
namely using the transition to the $^{13}$N($3/2^{-}, T=3/2)$ state at 15.1 MeV (labeled 3 in Fig.
\ref{levels}). This transition has three analogs, labeled 1,2 and 4 in Fig. \ref{levels}. For
transition 1 ($^{13}$B($\beta^{-}$)$^{13}$C) and transition 4 ($^{13}$O($\beta^{-}$)$^{13}$N) the
lifetimes, and thus B(GT) values, are known from $\beta$-decay experiments. After correcting for
the difference in spin and isospin factors \footnote{B(GT) is calculated from the matrix element
$M(GT)$ via $B(GT)=\frac{1}{2(2J_{i}+1)}\frac{\langle T_{i} T_{zi} \Delta T \Delta T_{z} | T_{f}
T_{zf}\rangle^{2}}{2T_{f}+1} M^2(GT)$}, the B(GT) for transition 3 is estimated to be
$0.213\pm0.001$ using the B(GT) from transition 4 and $0.237\pm0.001$ using the B(GT) from
transition 1. The mirror asymmetry in the GT $\beta$-decay matrix elements can be attributed to
isospin breaking due to the Coulomb interaction. In Ref. \cite{TOW73}, the ratio
$R=\frac{M^{2}(^{13}O(3/2^{-},g.s.)\rightarrow ^{13}N(1/2^{-}, g.s.))}{M^{2}(^{13}B(3/2^{-},
g.s.)\rightarrow ^{13}C(1/2^{-},g.s.))}$ was estimated to be 0.925 for the radial overlap (Method I
of Table 1 from Ref. \cite{TOW73}). Combined with value for isospin mixing in the $p$-shell for
this transition of 0.992, $R$ becomes 0.918, which is close to the measured asymmetry of
$\frac{0.213\pm0.001}{0.237\pm0.001}=0.899\pm0.006$. In Ref. \cite{TAD87}, the B(GT) for the
transition 3 of Fig. \ref{levels} was set to 0.23$\pm$0.01, where the error was chosen to represent
the minor ambiguity due to isospin symmetry breaking and, for the sake of consistency, we used the
same number in the current analysis.

We used the methods of Ref. \cite{TOW73} to estimate the isospin symmetry breaking between
transitions 5 and 6 in Fig. \ref{levels}. The calculations are based on the Skyrme Hartree-Fock
radial wave functions obtained with the Skx interaction \cite{BRO98}. The $p$-shell isospin
breaking was obtained in proton-neutron formalism with the charge-dependent interaction of Ormand
and Brown \cite{ORM89}. The result for $R=\frac{M^{2}(^{13}N(1/2^{-},g.s.)\rightarrow
^{13}C(3/2^{-},3.68 \text{ MeV}))}{M^{2}(^{13}C(1/2^{-},g.s.)\rightarrow ^{13}N(3/2^{-},3.51 \text{
MeV}))}=1.043$ for the radial overlap and 1.014 for the $p$-shell isospin mixing for this
transition for a total ratio  $R=1.06$. Since this estimated asymmetry is comparable in magnitude
to the uncertainties in extracting the B(GT) value for transition 6 via the charge-exchange
reactions, isospin symmetry is assumed in the remainder of this paper.

A minor disadvantage of using transition 3 to calibrate the GT unit cross section is that the
Q-value (17.248 MeV), and thus the minimum momentum transfer (at $0^{\circ}$) for this transition
are relatively large (0.18 fm$^{-1}$). Since the cross section at $q=0$ must be used in Eq.
\ref{eq:eik} an extrapolation is required which is estimated theoretically (see below). However,
the error in this extrapolation is small compared to the error in determining the GT unit cross
section from the ground-state transition where the Fermi component has to be subtracted.

After the determination of the GT unit cross section using transition 3, the B(GT)
for transition 6 can be extracted. Moreover, since the B(GT) for the ground-state
transition is known, the cross section for the GT component can be estimated and
subtracted from the total cross section for the ground-state excitation. The
remaining Fermi cross section is then used to determine the Fermi unit cross
section.

\section{Experimental results.}
\label{sec:data}

The $^{13}$C($^{3}$He,$t$) data were obtained at RCNP using the spectrometer Grand Raiden
\cite{FUJ99}. A 3-pnA, 420 MeV $^{3}$He beam bombarded a 4.1 mg/cm$^2$ $^{nat}$C target and a 1.0
mg/cm$^{2}$ $^{nat}$C$_{8}$H$_{8}$O$_{4}$ (Mylar) target. Tritons were measured in the focal plane
of the spectrometer and momenta and angles reconstructed through a ray-tracing procedure. The
spectrometer was tuned to run in the ``off-focus'' mode \cite{FUJ01} of operation, so that the
angles in both dispersive and non-dispersive direction could be extracted with good accuracy.
Center-of-mass scattering angles between 0$^{\circ}$ and 3.0$^{\circ}$ were covered with a
resolution of 0.2$^{\circ}$ (FWHM). The energy resolution was $\sim 110$ keV with the thin Mylar
target and $\sim 150$ keV with the $^{nat}$C target due to larger differential energy loss of the
$^{3}$He and triton in the thicker target. $^{3}$He beam particles were collected for the purpose
of cross section calculation in a Faraday cup placed in the inner bend of the first dipole magnet
of Grand Raiden. Since the cross sections for the two different targets were consistent within
statistical errors and the total number of events collected a factor of ten larger with the
$^{nat}$C target, the results reported here are those taken with the $^{nat}$C target. In the same
beam-time period as these data were taken, data for other targets were obtained
 to establish an empirical relationship for unit cross sections as a function
of mass number. The results of that study were presented in Ref. \cite{ZEG07}.

\begin{figure}
\includegraphics[width=7.5cm]{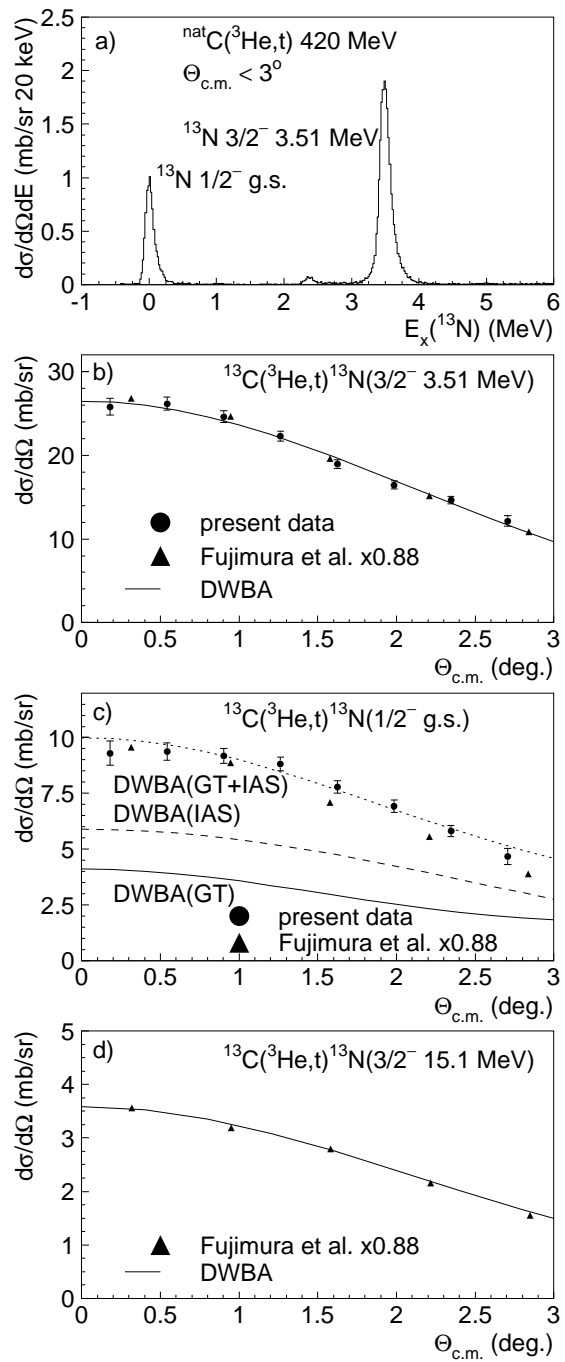}
\caption{\label{spectra} a) Excitation energy spectrum of $^{13}$N taken with the ($^{3}$He,$t$)
reaction at 420 MeV on a $^{nat}$C target. The ground state and excited state at 3.51 MeV are
indicated. b) Differential cross sections for the $^{13}$C($^{3}$He,$t$)$^{13}$N(3.51 \text{ MeV})
reaction from the present data (error bars are statistical only) and from Ref. \cite{FUJ04}. The
latter have been scaled by a factor 0.88. Also included is the result from the DWBA calculation,
for which the scale was adjusted to fit the data. c) Idem, but for the
$^{13}$C($^{3}$He,$t$)$^{13}$N(\text{g.s.}) transition. The contribution from GT and Fermi
components to this transition is described in the text. d) Idem, but for the
$^{13}$C($^{3}$He,$t$)$^{13}$N(15.1 \text{ MeV}) transition. Only data from Ref. \cite{FUJ04} is
available.}
\end{figure}

In Fig. \ref{spectra}a the excitation-energy spectrum of $^{13}$N taken with the
$^{nat}$C is shown for center-of-mass scattering angles between $0^{\circ}$ and
$3^{\circ}$. Besides the ground state and the state at 3.51 MeV, the $1/2^{+}$ state
at 2.36 MeV can also be seen. Fig. \ref{spectra}b shows the differential cross
sections for the transition to the 3.51 MeV state. The data from Ref. \cite{FUJ04}
are also shown after scaling by a factor of 0.88. The effect of the slight
difference in beam energy (420 MeV for the present data and 450 MeV for Ref.
\cite{FUJ04}) on the absolute cross section was estimated in DWBA (detailed below)
and was found responsible for about half of the required adjustment. The need for
further scaling is most likely due to systematic errors in the beam integrations and
target thicknesses. \footnote{Although the accuracies of beam integration and target thickness are not necessarily better in the newer data than in Ref. \cite{FUJ04}, we chose to scale the older cross sections to the newer ones to maintain consistency with Ref. \cite{ZEG07}, rather than taking the average of the two data sets.}.
Note that another state ($J^{\pi}=5/2^{+}$) exists at an
excitation energy of 3.55 MeV, which cannot be separated from the state at 3.51 MeV.
The transition to the 3.55 MeV state is of dipole nature. If it had a significant
cross section it would have drastically changed the angular distribution shown in
Fig. \ref{spectra}b, since dipole transitions have a minimum and maximum at
scattering angles of $0^{\circ}$ and $\sim 1.2^{\circ}$, respectively. We,
therefore, concluded that the contribution from this transition can be ignored.

Fig. \ref{spectra}c shows the measured differential cross sections to the $^{13}$N
ground state from both the present data and from Ref. \cite{FUJ04}. The latter has
been scaled by the same factor 0.88 described above. A slight discrepancy can be
seen between the data sets, which is likely caused by a reduction of the relative
contribution from the Fermi component to the excitation of this state at the higher
beam energy. The strength of the central $\tau$ interaction drops significantly as a
function of beam energy \cite{LOV81,LOV85} and is apparently noticeable even when
the beam energy changes only by 7\%, as is the case for the comparison between the
data sets at 420 MeV and 450 MeV.

Finally, in Fig. \ref{spectra}d, the differential cross section for the transition
to the $^{13}$N state at 15.1 MeV is shown after scaling by the factor 0.88. Only
cross sections from Ref. \cite{FUJ04} are available in this case because of the
contamination of $^{12}$C in the targets used for the later experiment (see above).
It should be noted that the $^{13}$C target used in Ref. \cite{FUJ04} was enriched
to at least 99\% and that the contamination from the $^{12}$N ground state into the
$^{13}$N excited state at 15.1 MeV is no more than 4\% based on the cross section
measured for the transition to the $^{12}$N ground state in the data obtained with
the $^{nat}$C target.

To determine the cross section at $q=0$, first the cross section at $0^{\circ}$ was obtained from
the measured differential cross sections shown in Figs. \ref{spectra}b-d. For the excitation of the
states at 3.51 MeV and 15.1 MeV, this was done by fitting the differential cross sections
calculated in distorted-wave Born approximation (DWBA) to the data using a single scaling factor.
The DWBA cross section at $0^{\circ}$ multiplied by the fitted scaling factor was then used as the
estimate for the experimental cross section at $0{^\circ}$ and its uncertainty determined from the
error in the fit.

The DWBA calculations were performed using the code FOLD \cite{FOLD}. in this code, the Love-Franey
nucleon-nucleon interaction \cite{LOV81,LOV85} is double-folded over the projectile-ejectile and
target-residue transition densities. Because the $^{3}$He and triton are composite particles, a
short-range approximation as described in Ref. \cite{LOV81} was used for the exchange terms in the
potential. For $^{3}$He and $^{3}$H, densities used in the folding were obtained from Variational
Monte-Carlo results \cite{WIR05}. One-body transition densities (OBTDs) for the transitions from
$^{13}$C to $^{13}$N were calculated with the code OXBASH \cite{OXBA} employing the CKII
interaction \cite{COH67} in the p shell-model space. Radial wave functions were calculated using a
Woods-Saxon potential. Binding energies of the particles were determined in OXBASH \cite{OXBA}
using the Skyrme SK20 interaction \cite{BRO98}. It should be noted that the B(GT) values for the
ground, 3.51-MeV and 15.1-MeV states, calculated with the CKII interaction, were
 0.190, 1.50 and 0.228,
respectively, after correcting for the quenching factor of 0.67. These are slightly
different, but a little closer to the experimental values from $\beta$-decay for the
ground-state and 15.1 MeV excitations, from those calculated with the WBT
interaction \cite{WAR92} in Ref. \cite{WAN01} (B(GT) values of 0.1746, 1.3444 and
0.2849, respectively).

Optical potential parameters extracted from $^{3}$He elastic scattering on $^{13}$C \cite{FUJ04}
were used in the DWBA calculation. Following Ref. \cite{WER89}, the depths of the triton potentials
were calculated by multiplying the depths of the $^{3}$He potentials by 0.85, while leaving radii
and diffusenesses constant.

For GT transitions, in principle a fit in which the $\Delta L=0$ and incoherent
$\Delta L=2$ pieces are independently scaled is required. However, inclusion of an
incoherent $\Delta L=2$ component changed the contribution from the $\Delta L=0$
component to the forward-angle cross section by less than 1\%. Since this change is
smaller than uncertainties due to statistical error margins,  the incoherent $\Delta
L=2$ contributions are not shown in Figs. \ref{spectra}b-d. Such small contributions
from the incoherent $\Delta L=2$ components are consistent with the DWBA
calculations.

For the transition to the ground state the procedure is complicated by the contribution from both
Fermi and GT components whose angular distributions at forward angles are very similar. The
contribution from the GT component must, therefore, first be fixed using the unit cross section
determined from the transition to the excited state at 15.1 MeV.

In Table \ref{tab:table1} the results from the extraction procedure are summarized. The B(GT)
values derived from the log$ft$ values obtained from $\beta$-decay are given in row a), where
available. For comparison, the B(GT) values calculated in the shell-model are shown in row b).
These strengths have been multiplied with the GT quenching factor of 0.67 \cite{CHO93}. A good
consistency with the values extracted from $\beta$-decay data is found for the transitions to the
ground state and the excited state at 15.1 MeV. In row c) the extracted differential cross sections at $0^{\circ}$ obtained from
the data as described above are displayed. The cross sections at $q=0$ are shown in row d). These
were obtained from the cross sections at $0^{\circ}$ by multiplying them with the ratio
$\left[\cfrac{d\sigma}{d\Omega}(q=0)/\cfrac{d\sigma}{d\Omega}(0^{\circ})\right]$. The differential
cross sections in the numerator and the denominator are both calculated in DWBA. These ratios are
1.01, 1.03 and 1.25 for the transitions to the ground state (the ratio is the same for the Fermi
and GT components), the excited state at 3.51 MeV and the excited state at 15.1 MeV, respectively.
The GT unit cross section (shown in row e) is calculated by dividing the cross section at $q=0$ in
row d) for the excited state at 15.1 MeV by the corresponding B(GT) shown in row a). This GT unit
cross section is then used to calculate the B(GT) for the excited state at 3.51 MeV, given in row
f). This B(GT) is close to the corresponding shell-model value provided in row b). The error bar in the experimental B(GT) value is due to statistical uncertainties in the data, uncertainties in the fitting procedure to the DWBA calculations and also includes the uncertainty in the unit cross section derived from the 15.1 MeV state as quoted in row e).

Finally, the Fermi unit cross section was derived from the transition to the ground
state. To do so, the estimated GT contribution in row g) to this excitation must be
subtracted first and this is done by multiplying the GT unit cross in row e) with
the B(GT) extracted from $\beta$-decay (given in row a) for the ground-state
transition. The remaining cross section, placed in row h) was then attributed to the
Fermi transition. Since B(F)=1, this cross section is also the Fermi unit cross
section, as shown in row i).

\begin{table*}
\caption{\label{tab:table1} Summary of the analysis for the ground state and excited states at 3.51
MeV and 15.1 MeV in $^{13}$N excited via the $^{13}$C($^{3}$He,$t$) reaction. Note that the error
margins in the experimental differential cross sections are due to statistical uncertainties in the
fitting of measured angular distributions only. Detailed descriptions of each of the rows, labeled
a)-i) in the table, are given in the text.}
\begin{ruledtabular}
\begin{tabular}{llccc}
 & & \multicolumn{3}{c} {state in $^{13}$N}  \\
\cline{3-5} \\
       &                                    & $1/2^{-}$ g.s. & $3/2^{-}$ 3.51 MeV & $3/2^{-}$ 15.1 MeV \\
\hline \\
a)&B(GT)$_{\beta-\text{decay}}$                & $0.207\pm0.002$ & -                  & $0.23\pm0.01$      \\
b)&B(GT)$_{\text{CKII}}\times0.67\footnotemark[1]$                    & 0.19            & 1.50               & 0.23               \\
c)&$\left[\frac{d\sigma}{d\Omega}(0^{\circ})\right]_{exp}$ (mb/sr)
                                            & $10.0\pm0.5$    & $26.5\pm0.3$       & $3.6\pm0.1$\\
d)&$\left[\frac{d\sigma}{d\Omega}(q=0)\right]_{exp}$ (mb/sr)
                                            & $10.1\pm0.5$    & $27.3\pm0.3$       & $4.5\pm0.1$\\
e)&$\hat{\sigma}_{\sigma\tau}$  (mb/sr)                & -               & -                  & $20\pm1$\\
f)&B(GT)                       & -               & $1.37\pm0.07$      & - \\
g)&$\left[\frac{d\sigma}{d\Omega}(q=0)\right]_{\Delta S=1}$ (mb/sr)
                                            & $4.2\pm0.2$\footnotemark[2]($5.0\pm0.2$\footnotemark[3])      & -                  & - \\
h)&$\left[\frac{d\sigma}{d\Omega}(q=0)\right]_{\Delta S=0}$ (mb/s r)
                                            & $5.9\pm0.5$\footnotemark[2]($5.1\pm0.5$\footnotemark[3])      & -                  & - \\
i)&$\hat{\sigma}_{\tau}$(mb/sr)\footnotemark[4]                & $5.9\pm0.5$\footnotemark[2]($5.1\pm0.5$\footnotemark[3])       & -                  & - \\
\end{tabular}
\end{ruledtabular}
\footnotetext[1]{The factor 0.67 represents the phenomenological quenching factor
for GT strength.} \footnotetext[2]{Without taking into account corrections due to
the tensor-$\tau$ interaction (for details, see text).} \footnotetext[3]{After
taking into account corrections due to the tensor-$\tau$ interaction (for details,
see text).} \footnotetext[4]{Calculated using $B(F)=(N-Z)=1$}.
\end{table*}

As mentioned above, the proportionality between B(GT) and differential cross section
at $q=0$ can be broken if the interference between the $\Delta L=0$ amplitude and
$\Delta L=2$ amplitude due to the non-central tensor-$\tau$ interaction is strong. This has been noted before in studies of GT strengths of relevance for astrophysical purposes (see e.g. Refs. \cite{HEG05,COL06}). Here, we follow the procedure previously used in Refs. \cite{ZEG06,COL06} to quantify the effects of such interference. Namely,
by comparing the differential cross sections calculated in DWBA with the full Love-Franey
interaction with those in which the tensor-$\tau$ component of the Love-Franey
interaction is removed. The relative changes between these two values were
comparable to statistical error bars for the transitions to the excited states at
3.51 and 15.1 MeV, but large for the GT component of the transition to the ground
state; including the tensor-$\tau$ interaction increases the cross section by 15\%.
A corrected estimate for the GT contribution to the cross section of the transition
to the ground state, which takes this increase into account, is given between
brackets in row g) of Table \ref{tab:table1}. Consequently, the estimate for the
component in the cross section for the transition to the ground state due to the
Fermi component decreases to the values shown between brackets in rows h). As a
result, the Fermi unit cross section also decreases to the value shown between
brackets in row i).

The extracted GT unit cross section of $20\pm1$ mb/sr agrees well with the empirical relationship
($\hat{\sigma}_{\sigma\tau}=109A^{-0.65}=20.6$ mb/sr for A=13) of unit cross section as a function
of mass number \cite{ZEG07}. The empirical relationship for the Fermi unit cross section
($\hat{\sigma}_{\tau}=72A^{-1.06}$) gives 4.75 mb/sr for A=13. This value is consistent with the
extracted Fermi cross section if the correction to the GT component for the transition to the
ground state due to the interference with the tensor-$\tau$ interaction is applied ($5.1\pm0.5$
mb/sr). Together with the previous studies in Refs. \cite{ZEG06,COL06}, this provides further
indication that the procedure for estimating the proportionality breaking described above is
reasonable.

The unit cross sections can also be calculated from theory, by dividing the GT and Fermi cross
sections calculated in DWBA by the shell-model B(GT) and B(F)=1, respectively. Values of
$\hat{\sigma}_{\sigma\tau}(\text{theory})=34.7$ mb/sr and $\hat{\sigma}_{\tau}(\text{theory})=6.31$
mb/sr are found, which are both significantly higher than the experimental values. Such an
overestimate of the DWBA cross section is seen over a wide target-mass range \cite{ZEG07}. Although
the reasons for this discrepancy are not well understood, a likely cause is the approximate
treatment of the exchange contributions which has been shown to lead to an overestimation of the
cross sections in the DWBA calculations \cite{UDA87,KIM00}. Other systematic discrepancies could be
due to ambiguities in the optical potential parameters employed in the DWBA calculation and
possible density-dependences of the effective nucleon-nucleon interaction.

\section{Comparison with $^{13}$C($p$,$n$) data}
In $^{13}$C($p$,$n$) experiments detailed in Refs. \cite{TAD82,GOO85,TAD87}, the GT
unit cross section was determined using measurements of the ground-state cross
section, combined with a measurement of the spin-flip probability $S_{NN}$. When the
extracted unit cross section was used to determine the B(GT) for the transition to
the excited state at 3.51 MeV, the afore-mentioned large discrepancy with the
shell-model calculations was found. It was furthermore noted that the GT unit cross
section derived from the transition to the 15.1 MeV state was consistent with that
determined using the ground-state transition and that the Fermi unit cross section
was exceptionally large.

In contrast to this, in Ref. \cite{WAT85}, it was suggested that the proportionality
between the cross section and GT strength was broken for the GT transition to the
$^{13}$N(g.s.). An effective ($p$,$n$) operator was introduced in the shell-model
calculations based on the description of quenching of GT strength in the $sd$-shell
\cite{BRO83}. This modified operator, designed to take into account medium effects,
contained a term $\delta_{p}(8\pi)^{1/2}[Y^{(2)}\otimes\mathbf{s}]^{\Delta
J=1}\tau_{\pm}$. This term mediates contributions from $\Delta L=2$ amplitudes that
interfere with the $\Delta L=0$ GT amplitudes. It was speculated \cite{WAT85} that
the apparent enhancement of $\delta_{p}$ in transitions going from $j=l-\frac{1}{2}$
to $j=l-\frac{1}{2}$ (such as the $^{13}$C(g.s.)$\rightarrow ^{13}$N(g.s.)) might be
due to $\Delta$-isobar admixtures in the reaction mechanism.

In Refs. \cite{MIL91,WAN01}, it was shown that the cross section measured to the 15.1 MeV state in
the earlier experiments \cite{TAD82,GOO85,TAD87} was too high, possibly due to contamination from
the $^{12}$C($p$,$n$)$^{12}$N(g.s.) transition. Furthermore, it was shown in Ref. \cite{MIL91} that
the cross sections measured for $^{13}$C($p$,$n$)$^{13}$N(15.1 MeV) reaction and its analog
$^{13}$C($n$,$p$)$^{13}$B(g.s.) were consistent, providing further indication that the use of the
former to calibrate the GT unit cross section is appropriate.

Following Ref. \cite{WAT85}, further arguments were provided in Ref. \cite{WAN01}
that the proportionality between GT strength and cross section for the ground-state
transition was broken, and that it should not be used to calibrate the GT unit cross
section to extract B(GT) values for the excitation of other states in $^{13}$N. The
authors of Ref. \cite{WAN01} quote a B(GT) for the transition to the excited state
at 3.51 MeV of $1.06\pm0.05$, based on the empirical unit cross sections obtained
with even-even mass targets in the ($p$,$n$) reaction \cite{TAD87}. However, they also determined a GT unit cross section using the excitation to the 15.1 MeV state.  Its
value was 22\% lower than that obtained using the empirical relationships. The
reduced value for the GT unit cross section would increase the B(GT) for the
transition to the excited state at 3.51 MeV to $1.29\pm0.06$, which is within error
bars consistent to the value extracted via the ($^{3}$He,$t$) reaction and close to
the shell-model value.

We should note that in both the ($p$,$n$) work discussed in Refs. \cite{WAT85,WAN01} and the
($^{3}$He,$t$) analysis described here a large breaking of the proportionality between B(GT) and
cross section for the ground state transition is reported. However, whereas in Refs.
\cite{WAT85,WAN01} these are modeled through the modified operator in the shell-model calculations
mentioned above, we find that the breaking in the ($^{3}$He,$t$) reaction at 420 MeV is a direct
result of the tensor-$\tau$ component in the effective nucleon-nucleon interaction. Since the
consequences of either description are similar (causing interference between $\Delta L=0$ and
$\Delta L=2$ amplitudes) they could well represent the same underlying mechanism.

We conclude that there is a significant spread in the B(GT) values reported for the transition to
the 3.51 MeV state in $^{13}$N, but that the most recent results from $^{13}$C($p$,$n$) and the new
results from $^{13}$C($^3$He,$t$) favor an empirical value much closer to the shell-model
calculations than the earlier ($p$,$n$) results. Nevertheless, for the purpose of studying the
effects on the pre-explosion evolutionary track of SNe Ia below, we used all available experimental
values, as well as the shell-model results.

\section{Electron-capture rates in the SN\MakeTextLowercase{e} I\MakeTextLowercase{a} environment.}
\label{sec:ecapture}

As mentioned above, the electron capture on $^{13}$N, populating states in $^{13}$C
can be investigated using the analog transitions from $^{13}$C to $^{13}$N. We
performed such a study. The log$ft$ for the ground state transition is known from
$\beta$-decay and thus fixed. Therefore, the only parameter that was varied in the
study was the B(GT) for the transition from the ground state of $^{13}$N to the
$3/2^{-}$ state at 3.68 MeV in $^{13}$C. As mentioned above, isospin symmetry was
assumed, so that the B(GT) values deduced from experiment using the transition from
the $^{13}$C ground state to the $3/2^{-}$ state at 3.51 MeV in $^{13}$N could be
applied directly. Electron captures onto \carbon[13] are blocked by the high
threshold for this reaction ($13.4\,\text{MeV}$).

\begin{figure}
\includegraphics[width=8.0cm]{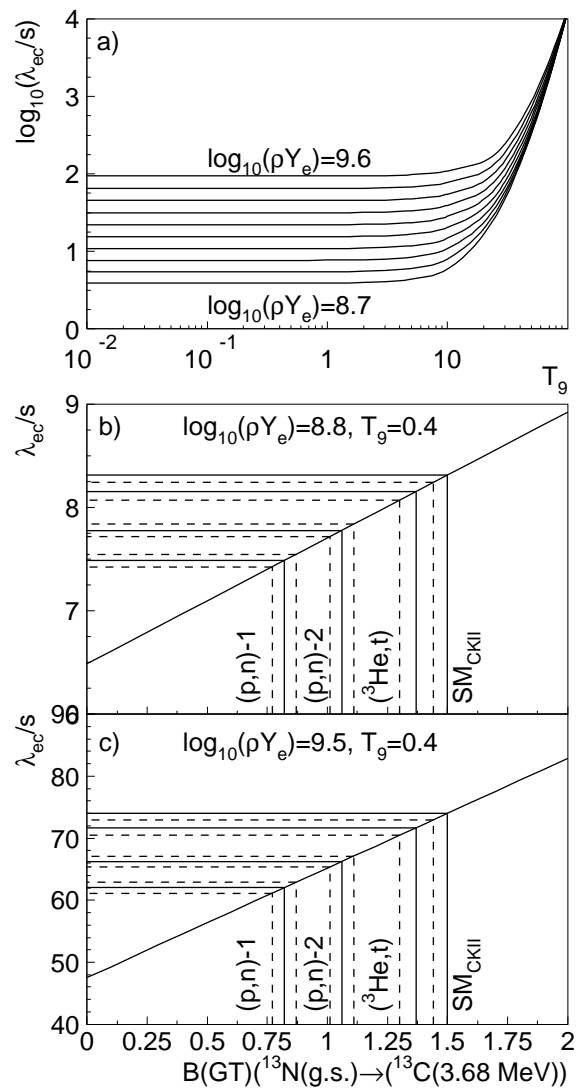}
\caption{\label{rates} a) Calculated electron-capture rates on $^{13}$N in the stellar environment
as a function of temperature and $\rho Y_{e}$ (ranging from log$_{10}$($\rho Y_{e}$)=8.7 to 9.6 in steps
of 0.1). The B(GT) for the transition to the 3.68 MeV excited state in $^{13}$C was fixed to the
shell-model value of 1.5. b) Electron-capture rate at log$_{10}$($\rho Y_{e}$)=8.8 and T$_{9}$=0.4 as a
function of the assumed B(GT) for the transition to the 3.68 MeV excited state in $^{13}$C. The
B(GT) values extracted from the ($p$,$n$) data (``1'' refers to Refs. \cite{TAD82,GOO85,TAD87} and
``2'' to Ref. \cite{WAN01}), ($^{3}$He,$t$) data and the shell-model calculation are indicated and
associated with electron-capture rates. The dashed lines correspond to error margins in the B(GT)
values deduced experimentally. c) idem, but at log$_{10}$($\rho Y_{e}$)=9.5.}
\end{figure}

The electron-capture rate calculations are based on the method described in Refs.\
\cite{fuller80,fuller82,fuller82a,fuller85}, which were implemented into a new code
\cite{BEC06}. The calculations were performed in a grid spanning log$_{10}$($\rho Y_{e}$)
values from 8.7 to 9.6 in steps of 0.1 and dimensionless temperatures
($T_{9}=T/10^{9}(\text{K})$) from 0.01 to 100. Electron chemical potentials
($\mu_{e}$) were computed from a tabulation \cite{timmes.swesty:accuracy}.  In Fig.
\ref{rates}a, the results of these calculations are shown for the case in which the
B(GT) for the transition to the excited state has been set to 1.5, which is the
shell-model value, corrected for quenching as used in Ref. \cite{CHA07}. Near the
temperature of relevance in the simmering stage of SNe Ia (approximately
$T_{9}=0.4$), the electron-capture rate is nearly independent of the temperature
since $kT$ is much smaller than the chemical potential $\mu_{e}$ and thus only
depends on the value of $\rho Y_{e}$.

The relevant $\log_{10}(\rho Y_{e})$ in the simmering stage ranges from about 8.8 to 9.5
and for the two extremes in this range, the electron-capture rates on $^{13}$N are
plotted as a function of the B(GT) for the transition to the excited state at 3.68
MeV in Figs.~\ref{rates}b and c. The temperature $T_9$ was fixed to 0.4 in both
cases. If the B(GT)=0 for the transition to the 3.68 MeV state, capture to the
$^{13}$C ground state is the only possible channel. From  Figs.~\ref{rates}b and c
one can thus read off the branching ratio for capture to the excited state as a
function of its B(GT). At $\log_{10}(\rho Y_{e})=8.8$, the electron-capture rate is then
6.49 s$^{-1}$. If the B(GT) for the transition to the first excited state is taken
from the older ($p$,$n$) data, the rate increases to 7.49$\pm$0.06 s$^{-1}$. Using
the newer ($p$,$n$) data, it increases to 7.78$\pm$0.06 s$^{-1}$. The rates
calculated based on the ($^{3}$He,$t$) data and the shell-model calculation are
close: 8.15$\pm$0.09 s$^{-1}$ and 8.31 s$^{-1}$. Including the transition to the
excited state increases the electron-capture rate, compared to that to the
ground-state transition only, by 15\%, 20\%, 26\%, and 28\% if the B(GT) values are
taken from the older ($p$,$n$) data, the latest ($p$,$n$) data, the ($^{3}$He,$t$)
data, and the shell-model calculation, respectively.

At log$_{10}$($\rho Y_{e})=9.5$, the electron-capture rate has increased to 47.5 s$^{-1}$ if the B(GT) of
the transition to the excited state is set to zero. The contributions from the transition to the
excited state increase that rate by an additional 30\%, 39\%, 51\%, and 56\% if the B(GT) values
are taken from the older ($p$,$n$) data, the newer ($p$,$n$) data, the ($^{3}$He,$t$) data, and the
shell-model calculation, respectively.

In a white dwarf star on the threshold of igniting and becoming a type Ia supernova, the timescale
for a sound wave to traverse the star is $t_{\mathrm{hyd}}\approx (G\langle\rho\rangle)^{-1/2}\sim
1\,\text{s}$, where $G$ is the gravitational constant and $\langle\rho\rangle$ is the average mass
density of the white dwarf. The electron capture rate is sufficiently fast, $\lambda_{\text{ec}} <
t_{\mathrm{hyd}}^{-1}$, that during the pre-explosion ``simmering phase,'' in which the heating from the
fusion of $\carbon + \carbon$ occurs on a timescale $>t_{\mathrm{hyd}}$, we should not expect the
increased electron-capture rate to affect the reaction flows that occur during this simmering
\cite{PIR07,CHA07}.  As mentioned in the introduction, captures into the excited state do affect, however, the heat deposited into the star from this reaction.  Since the \nitrogen[13] is produced from $\carbon(p,\gamma)\nitrogen[13]$ with the protons coming from the branch
$\carbon(\carbon,p)\sodium[23]$, the number of electron captures onto \nitrogen[13] is proportional to the number of \carbon\ consumed in raising the white dwarf temperature to ignition.  A greater heat release from $\nitrogen[13](e^{-},\nu_{e})\carbon[13]$ decreases the amount of carbon that must be consumed during this phase.

\begin{figure}
\includegraphics[width=8.0cm]{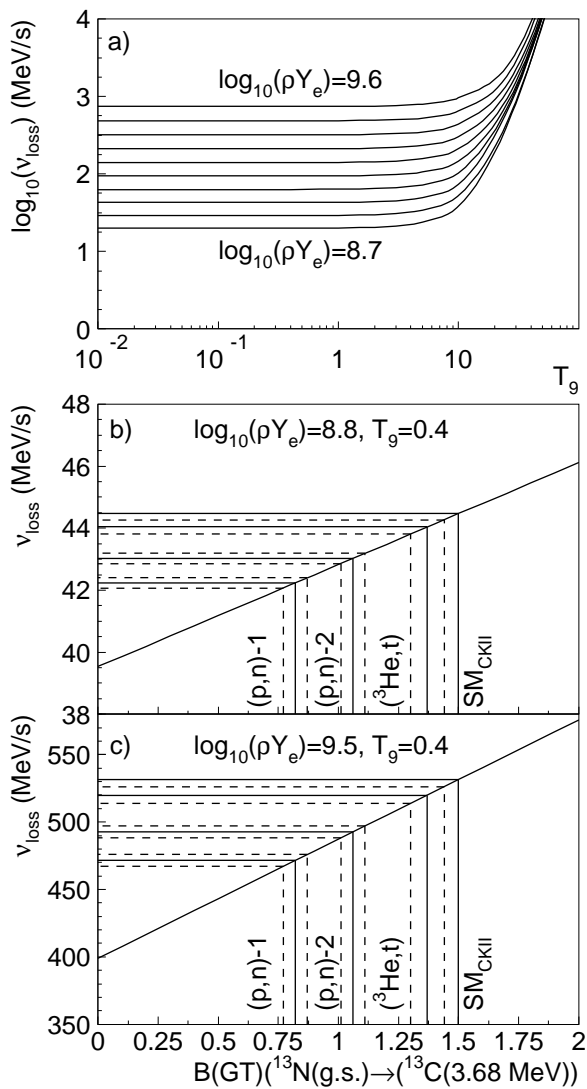}
\caption{\label{rates2}a) Calculated neutrino energy-loss rates due to the electron-capture on
$^{13}$N in the stellar environment as a function of temperature and $\rho Y_{e}$ (ranging from
log$_{10}$($\rho Y_{e}$)=8.7 to 9.6 in steps of 0.1). The B(GT) for the transition to the 3.68 MeV excited
state in $^{13}$C was fixed to the shell-model value of 1.5. b) neutrino energy-loss rate at
log$_{10}$($\rho Y_{e}$)=8.8 and T$_{9}$=0.4 as a function of the assumed B(GT) for the transition to the
3.68 MeV excited state in $^{13}$C. The B(GT) values extracted from the ($p$,$n$) data (``1''
refers to Refs. \cite{TAD82,GOO85,TAD87} and ``2'' to Ref. \cite{WAN01}), ($^{3}$He,$t$) data and
the shell-model calculation are indicated and associated with energy-loss rates. The dashed lines
correspond to error margins in the B(GT) values deduced experimentally. c) idem, but at log$_{10}$($\rho
Y_{e}$)=9.5. }
\end{figure}

The heat evolved from this reaction per second per \nitrogen[13] nucleus is
\begin{equation}\label{e.heat}
\dot{S} = \lambda_{\mathrm{ec}}\left(\mu_{e} + Q - \bar{E}_{\nu}\right),
\end{equation}
where the $Q$-value for the $\beta^{+}$-decay is 2.22 MeV, and $\bar{E}_{\nu}$ is
the mean energy of the neutrino.   We compute $\bar{E}_{\nu} =
\nu_{\mathrm{loss}}/\lambda_{\mathrm{ec}}$ from the integration over the electron
phase space, with $\lambda_{\mathrm{ec}}$ shown in Fig.~\ref{rates} and
$\nu_{\mathrm{loss}}$ shown in Fig.~\ref{rates2}. We find that, at $\log_{10}(\rho Y_{e})
= 8.8$, each reaction $\nitrogen[13](e^{-},\nu_{e})\carbon[13]$ deposits into the
white dwarf 0.53, 0.63, 0.77, and $0.81\,\text{MeV}$ for B(GT) taken from the older
$(p,n)$ data, the newer $(p,n)$ data, the $(\helium[3],t)$ data, and the shell-model
calculation, respectively. At a higher density, $\log_{10}(\rho Y_{e}) = 9.5$, the heat
deposition is 1.70, 1.89, 2.08, and $2.12\,\text{MeV}$ for these four cases.  If the
B(GT) were set to zero, the heat deposition from this reaction would be 0.07 and
$0.99\,\text{MeV}$ for $\log_{10}(\rho Y_{e})=8.8$ and 9.5, respectively.

We estimate the contribution of this reaction to the heating of the pre-explosive
white dwarf by integrating a reaction network at $\log_{10}(\rho Y_{e})=8.8$ and 9.5,
with the temperature at a fiducial value of $0.4\,\text{GK}$. The integration time
was chosen to be the heating timescale $C_{P}T/\dot{S}$, where $C_{P}$ is the
specific heat and $\dot{S}$ is the heating rate. The dominant heat sources are the reactions $\carbon+\carbon$, $\carbon(n,\gamma)\carbon[13]$,
$\carbon(p,\gamma)\nitrogen[13]$ and $\carbon[13](\alpha,n)\oxygen$, which release about $2.7\,\text{MeV}$ for each carbon consumed \cite{PIR07,CHA07}. At $\log_{10}(\rho Y_{e}) = 8.8$, the reaction
$\nitrogen[13](e^{-},\nu_{e})\carbon[13]$ contributes
0.4\%, 2.6\%, 3.1\%, 3.8\%, and 4.0\%
for $\mathrm{B(GT)}=0$, from the older $(p,n)$ data, the newer $(p,n)$
data, from the $(\helium[3],t)$ data, and from the shell-model calculation,
respectively. At $\log_{10}(\rho Y_{e}) = 9.5$, the higher electron Fermi energy ensures
that this reaction contributes somewhat more to the heating, although the
contribution is offset by a decrease in the production of \nitrogen[13].  We find
the contribution to the heating to be
4.3\%, 7.0\%, 7.8\%, 8.5\%, and 8.7\% for
$\mathrm{B(GT)}=0$, from the older $(p,n)$ data, from the newer $(p,n)$ data, from
the $(\helium[3],t)$ data, and from the shell-model calculation, respectively.
Although the differences between the results from the $(\helium[3],t)$ and the
$(p,n)$ data significantly affect the electron-capture rate and the net heat
deposition in the star due to this transition, we find that the overall effect on
the pre-explosion evolution of the white dwarf is small. The capture rate is
sufficiently fast in all cases that the rate of neutronization of the white dwarf is
controlled by the production of \nitrogen[13], and at relevant densities the heating
from this transition is a small component of the total.

\section{Conclusions}
\label{sec:conclusions} In summary, we extracted the GT strength for the transition from $^{13}$C
to the $^{13}$N 3/2$^{-}$ excited state at 3.51 MeV using the ($^{3}$He,$t$) reaction at 420 MeV
and existing data for the same reaction at 450 MeV. We find that the B(GT) is nearly consistent
with shell-model calculations using the CKII interaction and is higher than earlier experimental
values extracted using the $^{13}$C($p$,$n$) reaction. However, the discrepancies between
($^{3}$He,$t$) and ($p$,$n$) results are lifted if, in the analysis of latest ($p$,$n$)
experiments, the GT unit cross section is calibrated using the transition to the $3/2^{-}$ state at
15.1 MeV, as was done for the ($^{3}$He,$t$) reaction.

The Gamow-Teller component in the $^{13}$C($^{3}$He,$t$)$^{13}$N(g.s.) reaction was
estimated to be strongly increased because of interference between $\Delta L=0$ and
$\Delta L=2$ amplitudes. As a result, the estimated Fermi component for this
transition decreases. After this correction, both Fermi and GT unit cross sections
extracted from the data are consistent with the mass-dependent trends of Fermi and
GT unit cross sections, respectively, measured in even-even nuclei.

Under the assumption of isospin symmetry, the B(GT) for this transition is the same for the
transition from the $^{13}$N(g.s.) to the 3/2$^{-}$ state at 3.68 MeV in $^{13}$C. Combined with
the known weak transition strength from the $\beta$-decay of $^{13}$N to the ground state of
$^{13}$C, we investigated the impact of the differences between the extracted B(GT) values from the
($p$,$n$) data, the ($^{3}$He,$t$) data and the shell-model calculation on the evolution of Type Ia
supernovae during the pre-explosion simmering stage. We find that, although electron-capture and
neutrino energy-loss rates for this particular channel are significantly lower when the ($p$,$n$)
values are adopted instead of the ($^{3}$He,$t$) or shell-model values, the overall effect on the
evolution is small, since the heating due to this reaction is small compared to that from other
reactions. Nevertheless, the details of this reaction are now quantified for use in further
studies of SNe Ia.

\begin{acknowledgments}
We thank the cyclotron staff at RCNP for their support during the experiment described in this
paper and Dr. Fujimura for providing the detailed results of the earlier $^{13}$C($^{3}$He,$t$)
experiment described in Ref. \cite{FUJ04}. This work was supported by the US NSF (PHY0216783
(JINA), PHY-0555366, PHY-0606007, AST-0507456), the Ministry of Education, Science, Sports and
Culture of Japan and the Stichting voor Fundamenteel Onderzoek der Materie (FOM), the Netherlands.
\end{acknowledgments}

\bibliography{prc}% Produces the bibliography via BibTeX.

\begin{thebibliography}{47}
\expandafter\ifx\csname natexlab\endcsname\relax\def\natexlab#1{#1}\fi
\expandafter\ifx\csname bibnamefont\endcsname\relax
  \def\bibnamefont#1{#1}\fi
\expandafter\ifx\csname bibfnamefont\endcsname\relax
  \def\bibfnamefont#1{#1}\fi
\expandafter\ifx\csname citenamefont\endcsname\relax
  \def\citenamefont#1{#1}\fi
\expandafter\ifx\csname url\endcsname\relax
  \def\url#1{\texttt{#1}}\fi
\expandafter\ifx\csname urlprefix\endcsname\relax\def\urlprefix{URL }\fi
\providecommand{\bibinfo}[2]{#2}
\providecommand{\eprint}[2][]{\url{#2}}

\bibitem[{\citenamefont{Hillebrandt and Niemeyer}(2000)}]{HIL00}
\bibinfo{author}{\bibfnamefont{W.}~\bibnamefont{Hillebrandt}} \bibnamefont{and}
  \bibinfo{author}{\bibfnamefont{J.~C.} \bibnamefont{Niemeyer}},
  \bibinfo{journal}{ARA\&A} \textbf{\bibinfo{volume}{\bf{38}}},
  \bibinfo{pages}{191} (\bibinfo{year}{2000}).

\bibitem[{\citenamefont{{Sullivan} et~al.}(2006)\citenamefont{{Sullivan}, {Le
  Borgne}, {Pritchet}, {Hodsman}, {Neill}, {Howell}, {Carlberg}, {Astier},
  {Aubourg}, {Balam} et~al.}}]{Sullivan2006Rates-and-prope}
\bibinfo{author}{\bibfnamefont{M.}~\bibnamefont{{Sullivan}}},
  \bibinfo{author}{\bibfnamefont{D.}~\bibnamefont{{Le Borgne}}},
  \bibinfo{author}{\bibfnamefont{C.~J.} \bibnamefont{{Pritchet}}},
  \bibinfo{author}{\bibfnamefont{A.}~\bibnamefont{{Hodsman}}},
  \bibinfo{author}{\bibfnamefont{J.~D.} \bibnamefont{{Neill}}},
  \bibinfo{author}{\bibfnamefont{D.~A.} \bibnamefont{{Howell}}},
  \bibinfo{author}{\bibfnamefont{R.~G.} \bibnamefont{{Carlberg}}},
  \bibinfo{author}{\bibfnamefont{P.}~\bibnamefont{{Astier}}},
  \bibinfo{author}{\bibfnamefont{E.}~\bibnamefont{{Aubourg}}},
  \bibinfo{author}{\bibfnamefont{D.}~\bibnamefont{{Balam}}},
  \bibnamefont{et~al.}, \bibinfo{journal}{\apj} \textbf{\bibinfo{volume}{648}},
  \bibinfo{pages}{868} (\bibinfo{year}{2006}).

\bibitem[{\citenamefont{Brachwitz et~al.}(2000)\citenamefont{Brachwitz, Dean,
  Hix, Iwamoto, Langanke, Mart\'{i}nez-Pinedo, Nomoto, Strayer, Thielemann, and
  Umeda}}]{BRA00}
\bibinfo{author}{\bibfnamefont{F.}~\bibnamefont{Brachwitz}},
  \bibinfo{author}{\bibfnamefont{D.}~\bibnamefont{Dean}},
  \bibinfo{author}{\bibfnamefont{W.}~\bibnamefont{Hix}},
  \bibinfo{author}{\bibfnamefont{K.}~\bibnamefont{Iwamoto}},
  \bibinfo{author}{\bibfnamefont{K.}~\bibnamefont{Langanke}},
  \bibinfo{author}{\bibfnamefont{G.}~\bibnamefont{Mart\'{i}nez-Pinedo}},
  \bibinfo{author}{\bibfnamefont{K.}~\bibnamefont{Nomoto}},
  \bibinfo{author}{\bibfnamefont{M.~R.} \bibnamefont{Strayer}},
  \bibinfo{author}{\bibfnamefont{F.-K.} \bibnamefont{Thielemann}},
  \bibnamefont{and} \bibinfo{author}{\bibfnamefont{H.}~\bibnamefont{Umeda}},
  \bibinfo{journal}{Astrophys. J.} \textbf{\bibinfo{volume}{\bf{536}}},
  \bibinfo{pages}{934} (\bibinfo{year}{2000}).

\bibitem[{\citenamefont{{Timmes} et~al.}(2003)\citenamefont{{Timmes}, {Brown},
  and {Truran}}}]{timmes.brown.ea:variations}
\bibinfo{author}{\bibfnamefont{F.~X.} \bibnamefont{{Timmes}}},
  \bibinfo{author}{\bibfnamefont{E.~F.} \bibnamefont{{Brown}}},
  \bibnamefont{and} \bibinfo{author}{\bibfnamefont{J.~W.}
  \bibnamefont{{Truran}}}, \bibinfo{journal}{Astrophys. J.}
  \textbf{\bibinfo{volume}{\textbf{590}}}, \bibinfo{pages}{L83}
  (\bibinfo{year}{2003}).

\bibitem[{\citenamefont{Chamulak et~al.}(2008)\citenamefont{Chamulak, Brown,
  Timmes, and Dupczak}}]{CHA07}
\bibinfo{author}{\bibfnamefont{D.~A.} \bibnamefont{Chamulak}},
  \bibinfo{author}{\bibfnamefont{E.~F.} \bibnamefont{Brown}},
  \bibinfo{author}{\bibfnamefont{F.~X.} \bibnamefont{Timmes}},
  \bibnamefont{and} \bibinfo{author}{\bibfnamefont{K.}~\bibnamefont{Dupczak}},
  \bibinfo{journal}{Astrophys. J.}  (\bibinfo{year}{2008}), \bibinfo{note}{in
  press.}

\bibitem[{\citenamefont{Piro and Bildsten}(2008)}]{PIR07}
\bibinfo{author}{\bibfnamefont{A.~L.} \bibnamefont{Piro}} \bibnamefont{and}
  \bibinfo{author}{\bibfnamefont{L.}~\bibnamefont{Bildsten}},
  \bibinfo{journal}{Astrophys. J.}  (\bibinfo{year}{2008}), \bibinfo{note}{in
  press.}

\bibitem[{\citenamefont{Woosley et~al.}(2004)\citenamefont{Woosley, Wunsch, and
  Kuhlen}}]{Woosley2004Carbon-Ignition}
\bibinfo{author}{\bibfnamefont{S.~E.} \bibnamefont{Woosley}},
  \bibinfo{author}{\bibfnamefont{S.}~\bibnamefont{Wunsch}}, \bibnamefont{and}
  \bibinfo{author}{\bibfnamefont{M.}~\bibnamefont{Kuhlen}},
  \bibinfo{journal}{Astrophys. J.} \textbf{\bibinfo{volume}{\bf{607}}},
  \bibinfo{pages}{921} (\bibinfo{year}{2004}).

\bibitem[{\citenamefont{Kuhlen et~al.}(2006)\citenamefont{Kuhlen, Woosley, and
  Glatzmaier}}]{Kuhlen2006Carbon-Ignition}
\bibinfo{author}{\bibfnamefont{M.}~\bibnamefont{Kuhlen}},
  \bibinfo{author}{\bibfnamefont{S.~E.} \bibnamefont{Woosley}},
  \bibnamefont{and} \bibinfo{author}{\bibfnamefont{G.~A.}
  \bibnamefont{Glatzmaier}}, \bibinfo{journal}{Astrophys. J.}
  \textbf{\bibinfo{volume}{\bf{640}}}, \bibinfo{pages}{407}
  (\bibinfo{year}{2006}).

\bibitem[{\citenamefont{Ajzenberg-Selove}(1991)}]{AJZ91}
\bibinfo{author}{\bibfnamefont{F.}~\bibnamefont{Ajzenberg-Selove}},
  \bibinfo{journal}{Nucl. Phys.} \textbf{\bibinfo{volume}{\bf{A523}}},
  \bibinfo{pages}{1} (\bibinfo{year}{1991}).

\bibitem[{\citenamefont{Brown{\it{ et al.}}}()}]{OXBA}
\bibinfo{author}{\bibfnamefont{B.~A.} \bibnamefont{Brown{\it{ et al.}}}},
  \bibinfo{note}{{N}SCL report MSUCL-1289}.

\bibitem[{\citenamefont{Cohen and Kurath}(1967)}]{COH67}
\bibinfo{author}{\bibfnamefont{S.}~\bibnamefont{Cohen}} \bibnamefont{and}
  \bibinfo{author}{\bibfnamefont{D.}~\bibnamefont{Kurath}},
  \bibinfo{journal}{Nucl. Phys.} \textbf{\bibinfo{volume}{\bf{A101}}},
  \bibinfo{pages}{1} (\bibinfo{year}{1967}).

\bibitem[{\citenamefont{Chou et~al.}(1993)\citenamefont{Chou, Warburton, and
  Brown}}]{CHO93}
\bibinfo{author}{\bibfnamefont{W.-T.} \bibnamefont{Chou}},
  \bibinfo{author}{\bibfnamefont{E.~K.} \bibnamefont{Warburton}},
  \bibnamefont{and} \bibinfo{author}{\bibfnamefont{B.~A.} \bibnamefont{Brown}},
  \bibinfo{journal}{Phys. Rev. C} \textbf{\bibinfo{volume}{\bf{47}}},
  \bibinfo{pages}{163} (\bibinfo{year}{1993}).

\bibitem[{\citenamefont{Taddeucci{\it{ et al.}}}(1982)}]{TAD82}
\bibinfo{author}{\bibfnamefont{T.~N.} \bibnamefont{Taddeucci{\it{ et al.}}}},
  \bibinfo{journal}{Phys. Rev. C} \textbf{\bibinfo{volume}{\bf{25}}},
  \bibinfo{pages}{1094} (\bibinfo{year}{1982}).

\bibitem[{\citenamefont{Goodman{\it{ et al.}}}(1985)}]{GOO85}
\bibinfo{author}{\bibfnamefont{C.~D.} \bibnamefont{Goodman{\it{ et al.}}}},
  \bibinfo{journal}{Phys. Rev. Lett.} \textbf{\bibinfo{volume}{\bf{54}}},
  \bibinfo{pages}{877} (\bibinfo{year}{1985}).

\bibitem[{\citenamefont{Taddeucci{\it{ et al.}}}(1987)}]{TAD87}
\bibinfo{author}{\bibfnamefont{T.~D.} \bibnamefont{Taddeucci{\it{ et al.}}}},
  \bibinfo{journal}{Nucl. Phys.} \textbf{\bibinfo{volume}{\bf{A469}}},
  \bibinfo{pages}{125} (\bibinfo{year}{1987}).

\bibitem[{\citenamefont{Watson et~al.}(1985)\citenamefont{Watson, Pairsuwan,
  Anderson, Baldwin, Flanders, Madey, McCarthy, Brown, Wildenthal, and
  Foster}}]{WAT85}
\bibinfo{author}{\bibfnamefont{J.~W.} \bibnamefont{Watson}},
  \bibinfo{author}{\bibfnamefont{W.}~\bibnamefont{Pairsuwan}},
  \bibinfo{author}{\bibfnamefont{B.~D.} \bibnamefont{Anderson}},
  \bibinfo{author}{\bibfnamefont{A.~R.} \bibnamefont{Baldwin}},
  \bibinfo{author}{\bibfnamefont{B.~S.} \bibnamefont{Flanders}},
  \bibinfo{author}{\bibfnamefont{R.}~\bibnamefont{Madey}},
  \bibinfo{author}{\bibfnamefont{R.~J.} \bibnamefont{McCarthy}},
  \bibinfo{author}{\bibfnamefont{B.~A.} \bibnamefont{Brown}},
  \bibinfo{author}{\bibfnamefont{B.~H.} \bibnamefont{Wildenthal}},
  \bibnamefont{and} \bibinfo{author}{\bibfnamefont{C.~C.}
  \bibnamefont{Foster}}, \bibinfo{journal}{Phys. Rev. Lett.}
  \textbf{\bibinfo{volume}{55}}, \bibinfo{pages}{1369} (\bibinfo{year}{1985}).

\bibitem[{\citenamefont{Wang{\it{ et al.}}}(2001)}]{WAN01}
\bibinfo{author}{\bibfnamefont{X.}~\bibnamefont{Wang{\it{ et al.}}}},
  \bibinfo{journal}{Phys. Rev. C} \textbf{\bibinfo{volume}{63}},
  \bibinfo{pages}{024608} (\bibinfo{year}{2001}).

\bibitem[{\citenamefont{Fujimura{\it{ et al.}}}(2004)}]{FUJ04}
\bibinfo{author}{\bibfnamefont{H.}~\bibnamefont{Fujimura{\it{ et al.}}}},
  \bibinfo{journal}{Phys. Rev. C} \textbf{\bibinfo{volume}{\bf{69}}},
  \bibinfo{pages}{064327} (\bibinfo{year}{2004}), \bibinfo{note}{and private
  communication.}

\bibitem[{\citenamefont{Zegers{\it{ et al.}}}(2007)}]{ZEG07}
\bibinfo{author}{\bibfnamefont{R.~G.~T.} \bibnamefont{Zegers{\it{ et al.}}}},
  \bibinfo{journal}{Phys. Rev. Lett.} \textbf{\bibinfo{volume}{\bf{99}}},
  \bibinfo{pages}{202501} (\bibinfo{year}{2007}).

\bibitem[{\citenamefont{Zegers{\it{ et al.}}}(2006)}]{ZEG06}
\bibinfo{author}{\bibfnamefont{R.~G.~T.} \bibnamefont{Zegers{\it{ et al.}}}},
  \bibinfo{journal}{Phys. Rev. C} \textbf{\bibinfo{volume}{\bf{74}}},
  \bibinfo{pages}{024309} (\bibinfo{year}{2006}).

\bibitem[{\citenamefont{Cole{\it{ et al.}}}(2006)}]{COL06}
\bibinfo{author}{\bibfnamefont{A.~L.} \bibnamefont{Cole{\it{ et al.}}}},
  \bibinfo{journal}{Phys. Rev. C} \textbf{\bibinfo{volume}{\bf{74}}},
  \bibinfo{pages}{034333} (\bibinfo{year}{2006}).

\bibitem[{\citenamefont{Fujita{\it{ et al.}}}(2007)}]{FUJ07}
\bibinfo{author}{\bibfnamefont{Y.}~\bibnamefont{Fujita{\it{ et al.}}}},
  \bibinfo{journal}{Phys. Rev. C} \textbf{\bibinfo{volume}{\bf{75}}},
  \bibinfo{pages}{057305} (\bibinfo{year}{2007}).

\bibitem[{\citenamefont{Ormand and Brown}(1995)}]{ORM95}
\bibinfo{author}{\bibfnamefont{W.~E.} \bibnamefont{Ormand}} \bibnamefont{and}
  \bibinfo{author}{\bibfnamefont{B.~A.} \bibnamefont{Brown}},
  \bibinfo{journal}{Phys. Rev. C} \textbf{\bibinfo{volume}{\bf{52}}},
  \bibinfo{pages}{2455} (\bibinfo{year}{1995}).

\bibitem[{\citenamefont{Dubbers}(1991)}]{DUB91}
\bibinfo{author}{\bibfnamefont{D.}~\bibnamefont{Dubbers}},
  \bibinfo{journal}{Nucl. Phys.} \textbf{\bibinfo{volume}{\bf{A527}}},
  \bibinfo{pages}{527c} (\bibinfo{year}{1991}).

\bibitem[{\citenamefont{Hardy et~al.}(1990)\citenamefont{Hardy, Towner,
  Koslowsky, Hagberg, and Schmeing}}]{HAR90}
\bibinfo{author}{\bibfnamefont{J.~C.} \bibnamefont{Hardy}},
  \bibinfo{author}{\bibfnamefont{I.~S.} \bibnamefont{Towner}},
  \bibinfo{author}{\bibfnamefont{V.~T.} \bibnamefont{Koslowsky}},
  \bibinfo{author}{\bibfnamefont{E.}~\bibnamefont{Hagberg}}, \bibnamefont{and}
  \bibinfo{author}{\bibfnamefont{H.}~\bibnamefont{Schmeing}},
  \bibinfo{journal}{Nucl. Phys.} \textbf{\bibinfo{volume}{\bf{A509}}},
  \bibinfo{pages}{429} (\bibinfo{year}{1990}).

\bibitem[{\citenamefont{Towner}(1973)}]{TOW73}
\bibinfo{author}{\bibfnamefont{I.~S.} \bibnamefont{Towner}},
  \bibinfo{journal}{Nucl. Phys.} \textbf{\bibinfo{volume}{\bf{A216}}},
  \bibinfo{pages}{589} (\bibinfo{year}{1973}).

\bibitem[{\citenamefont{Brown}(1998)}]{BRO98}
\bibinfo{author}{\bibfnamefont{B.~A.} \bibnamefont{Brown}},
  \bibinfo{journal}{Phys. Rev. C} \textbf{\bibinfo{volume}{\bf{58}}},
  \bibinfo{pages}{220} (\bibinfo{year}{1998}).

\bibitem[{\citenamefont{Ormand and Brown}(1989)}]{ORM89}
\bibinfo{author}{\bibfnamefont{W.~E.} \bibnamefont{Ormand}} \bibnamefont{and}
  \bibinfo{author}{\bibfnamefont{B.~A.} \bibnamefont{Brown}},
  \bibinfo{journal}{Nucl. Phys.} \textbf{\bibinfo{volume}{\bf{A491}}},
  \bibinfo{pages}{1} (\bibinfo{year}{1989}).

\bibitem[{\citenamefont{Fujiwara{\it{ et al.}}}(1999)}]{FUJ99}
\bibinfo{author}{\bibfnamefont{M.}~\bibnamefont{Fujiwara{\it{ et al.}}}},
  \bibinfo{journal}{Nucl. Instrum. Meth. Phys. Res. A}
  \textbf{\bibinfo{volume}{\bf{422}}}, \bibinfo{pages}{484}
  (\bibinfo{year}{1999}).

\bibitem[{\citenamefont{Fujita{\it{ et al.}}}(2001)}]{FUJ01}
\bibinfo{author}{\bibfnamefont{H.}~\bibnamefont{Fujita{\it{ et al.}}}},
  \bibinfo{journal}{Nucl. Instrum. Meth. Phys. Res. A}
  \textbf{\bibinfo{volume}{\bf{469}}}, \bibinfo{pages}{55}
  (\bibinfo{year}{2001}).

\bibitem[{\citenamefont{Love and Franey}(1981)}]{LOV81}
\bibinfo{author}{\bibfnamefont{W.~G.} \bibnamefont{Love}} \bibnamefont{and}
  \bibinfo{author}{\bibfnamefont{M.~A.} \bibnamefont{Franey}},
  \bibinfo{journal}{Phys. Rev. C} \textbf{\bibinfo{volume}{\bf{24}}},
  \bibinfo{pages}{1073} (\bibinfo{year}{1981}).

\bibitem[{\citenamefont{Franey and Love}(1985)}]{LOV85}
\bibinfo{author}{\bibfnamefont{M.~A.} \bibnamefont{Franey}} \bibnamefont{and}
  \bibinfo{author}{\bibfnamefont{W.~G.} \bibnamefont{Love}},
  \bibinfo{journal}{Phys. Rev. C} \textbf{\bibinfo{volume}{\bf{31}}},
  \bibinfo{pages}{488} (\bibinfo{year}{1985}).

\bibitem[{\citenamefont{Cook and Carr}(1988)}]{FOLD}
\bibinfo{author}{\bibfnamefont{J.}~\bibnamefont{Cook}} \bibnamefont{and}
  \bibinfo{author}{\bibfnamefont{J.}~\bibnamefont{Carr}}
  (\bibinfo{year}{1988}), \bibinfo{note}{computer program \textsc{fold},
  Florida State University (unpublished), based on F. Petrovich and D. Stanley,
  Nucl. Phys. {\bf{A275}}, 487 (1977), modified as described in J. Cook {\it{
  et al.}}, Phys. Rev. C {\bf{30}}, 1538 (1984) and R. G. T. Zegers, S.
  Fracasso and G. Col\`{o} (2006), unpublished.}

\bibitem[{\citenamefont{Pieper and Wiringa}(2001)}]{WIR05}
\bibinfo{author}{\bibfnamefont{S.~C.} \bibnamefont{Pieper}} \bibnamefont{and}
  \bibinfo{author}{\bibfnamefont{R.~B.} \bibnamefont{Wiringa}},
  \bibinfo{journal}{Annu. Rev. Nucl. Part. Sci.}
  \textbf{\bibinfo{volume}{\bf{51}}}, \bibinfo{pages}{53}
  (\bibinfo{year}{2001}), \bibinfo{note}{and R.B. Wiringa, private
  communication}.

\bibitem[{\citenamefont{Warburton and Brown}(1992)}]{WAR92}
\bibinfo{author}{\bibfnamefont{E.~K.} \bibnamefont{Warburton}}
  \bibnamefont{and} \bibinfo{author}{\bibfnamefont{B.~A.} \bibnamefont{Brown}},
  \bibinfo{journal}{Phys. Rev. C} \textbf{\bibinfo{volume}{\bf{46}}},
  \bibinfo{pages}{923} (\bibinfo{year}{1992}).

\bibitem[{\citenamefont{van~der Werf et~al.}(1989)\citenamefont{van~der Werf,
  Brandenburg, Grasdijk, Sterrenburg, Harakeh, Greenfield, Brown, and
  Fujiwara}}]{WER89}
\bibinfo{author}{\bibfnamefont{S.~Y.} \bibnamefont{van~der Werf}},
  \bibinfo{author}{\bibfnamefont{S.}~\bibnamefont{Brandenburg}},
  \bibinfo{author}{\bibfnamefont{P.}~\bibnamefont{Grasdijk}},
  \bibinfo{author}{\bibfnamefont{W.~A.} \bibnamefont{Sterrenburg}},
  \bibinfo{author}{\bibfnamefont{M.~N.} \bibnamefont{Harakeh}},
  \bibinfo{author}{\bibfnamefont{M.~B.} \bibnamefont{Greenfield}},
  \bibinfo{author}{\bibfnamefont{B.~A.} \bibnamefont{Brown}}, \bibnamefont{and}
  \bibinfo{author}{\bibfnamefont{M.}~\bibnamefont{Fujiwara}},
  \bibinfo{journal}{Nucl. Phys.} \textbf{\bibinfo{volume}{\bf{A496}}},
  \bibinfo{pages}{305} (\bibinfo{year}{1989}).

\bibitem[{\citenamefont{Heger et~al.}(2005)\citenamefont{Heger, Kolbe, Haxton,
  Langanke, Mart\'{i}nez-Pinedo, and Woosley}}]{HEG05}
\bibinfo{author}{\bibfnamefont{A.}~\bibnamefont{Heger}},
  \bibinfo{author}{\bibfnamefont{E.}~\bibnamefont{Kolbe}},
  \bibinfo{author}{\bibfnamefont{W.~C.} \bibnamefont{Haxton}},
  \bibinfo{author}{\bibfnamefont{K.}~\bibnamefont{Langanke}},
  \bibinfo{author}{\bibfnamefont{G.}~\bibnamefont{Mart\'{i}nez-Pinedo}},
  \bibnamefont{and} \bibinfo{author}{\bibfnamefont{S.~E.}
  \bibnamefont{Woosley}}, \bibinfo{journal}{Phys. Lett.}
  \textbf{\bibinfo{volume}{\bf{606B}}}, \bibinfo{pages}{258}
  (\bibinfo{year}{2005}).

\bibitem[{\citenamefont{Udagawa et~al.}(1987)\citenamefont{Udagawa, Schulte,
  and Osterfeld}}]{UDA87}
\bibinfo{author}{\bibfnamefont{T.}~\bibnamefont{Udagawa}},
  \bibinfo{author}{\bibfnamefont{A.}~\bibnamefont{Schulte}}, \bibnamefont{and}
  \bibinfo{author}{\bibfnamefont{F.}~\bibnamefont{Osterfeld}},
  \bibinfo{journal}{Nucl. Phys.} \textbf{\bibinfo{volume}{\bf{A474}}},
  \bibinfo{pages}{131} (\bibinfo{year}{1987}).

\bibitem[{\citenamefont{Kim et~al.}(2000)\citenamefont{Kim, Knobles, Stotts,
  and Udagawa}}]{KIM00}
\bibinfo{author}{\bibfnamefont{B.~T.} \bibnamefont{Kim}},
  \bibinfo{author}{\bibfnamefont{D.~P.} \bibnamefont{Knobles}},
  \bibinfo{author}{\bibfnamefont{S.~A.} \bibnamefont{Stotts}},
  \bibnamefont{and} \bibinfo{author}{\bibfnamefont{T.}~\bibnamefont{Udagawa}},
  \bibinfo{journal}{Phys. Rev. C} \textbf{\bibinfo{volume}{\bf{61}}},
  \bibinfo{pages}{044611} (\bibinfo{year}{2000}).

\bibitem[{\citenamefont{Brown and Wildenthal}(1983)}]{BRO83}
\bibinfo{author}{\bibfnamefont{B.~A.} \bibnamefont{Brown}} \bibnamefont{and}
  \bibinfo{author}{\bibfnamefont{B.~H.} \bibnamefont{Wildenthal}},
  \bibinfo{journal}{Phys. Rev. C} \textbf{\bibinfo{volume}{\bf{28}}},
  \bibinfo{pages}{2397} (\bibinfo{year}{1983}).

\bibitem[{\citenamefont{Mildenberger et~al.}(1991)\citenamefont{Mildenberger,
  Alford, Celler, H{\"{a}}usser, Jackson, Larson, Pointon, Trudel, Vetterli,
  and Yen}}]{MIL91}
\bibinfo{author}{\bibfnamefont{J.~L.} \bibnamefont{Mildenberger}},
  \bibinfo{author}{\bibfnamefont{W.~P.} \bibnamefont{Alford}},
  \bibinfo{author}{\bibfnamefont{A.}~\bibnamefont{Celler}},
  \bibinfo{author}{\bibfnamefont{O.}~\bibnamefont{H{\"{a}}usser}},
  \bibinfo{author}{\bibfnamefont{K.~P.} \bibnamefont{Jackson}},
  \bibinfo{author}{\bibfnamefont{B.}~\bibnamefont{Larson}},
  \bibinfo{author}{\bibfnamefont{B.}~\bibnamefont{Pointon}},
  \bibinfo{author}{\bibfnamefont{A.}~\bibnamefont{Trudel}},
  \bibinfo{author}{\bibfnamefont{M.~C.} \bibnamefont{Vetterli}},
  \bibnamefont{and} \bibinfo{author}{\bibfnamefont{S.}~\bibnamefont{Yen}},
  \bibinfo{journal}{Phys. Rev. C} \textbf{\bibinfo{volume}{\bf{43}}},
  \bibinfo{pages}{1777} (\bibinfo{year}{1991}).

\bibitem[{\citenamefont{Fuller et~al.}(1980)\citenamefont{Fuller, Fowler, and
  Newman}}]{fuller80}
\bibinfo{author}{\bibfnamefont{G.~M.} \bibnamefont{Fuller}},
  \bibinfo{author}{\bibfnamefont{W.~A.} \bibnamefont{Fowler}},
  \bibnamefont{and} \bibinfo{author}{\bibfnamefont{M.~J.}
  \bibnamefont{Newman}}, \bibinfo{journal}{Astrophys. J. Supp. Series}
  \textbf{\bibinfo{volume}{42}}, \bibinfo{pages}{447} (\bibinfo{year}{1980}).

\bibitem[{\citenamefont{Fuller}(1982)}]{fuller82}
\bibinfo{author}{\bibfnamefont{G.~M.} \bibnamefont{Fuller}},
  \bibinfo{journal}{Astrophys. J.} \textbf{\bibinfo{volume}{252}},
  \bibinfo{pages}{741} (\bibinfo{year}{1982}).

\bibitem[{\citenamefont{Fuller et~al.}(1982)\citenamefont{Fuller, Fowler, and
  Newman}}]{fuller82a}
\bibinfo{author}{\bibfnamefont{G.~M.} \bibnamefont{Fuller}},
  \bibinfo{author}{\bibfnamefont{W.~A.} \bibnamefont{Fowler}},
  \bibnamefont{and} \bibinfo{author}{\bibfnamefont{M.~J.}
  \bibnamefont{Newman}}, \bibinfo{journal}{Astrophys. J.}
  \textbf{\bibinfo{volume}{252}}, \bibinfo{pages}{715} (\bibinfo{year}{1982}).

\bibitem[{\citenamefont{Fuller et~al.}(1985)\citenamefont{Fuller, Fowler, and
  Newman}}]{fuller85}
\bibinfo{author}{\bibfnamefont{G.~M.} \bibnamefont{Fuller}},
  \bibinfo{author}{\bibfnamefont{W.~A.} \bibnamefont{Fowler}},
  \bibnamefont{and} \bibinfo{author}{\bibfnamefont{M.~J.}
  \bibnamefont{Newman}}, \bibinfo{journal}{Astrophys. J.}
  \textbf{\bibinfo{volume}{293}}, \bibinfo{pages}{1} (\bibinfo{year}{1985}).

\bibitem[{\citenamefont{Becerril-Reyes
  et~al.}(2006)\citenamefont{Becerril-Reyes, Gupta, Schatz, Kratz, and
  M\"{o}ller}}]{BEC06}
\bibinfo{author}{\bibfnamefont{A.~D.} \bibnamefont{Becerril-Reyes}},
  \bibinfo{author}{\bibfnamefont{S.}~\bibnamefont{Gupta}},
  \bibinfo{author}{\bibfnamefont{H.}~\bibnamefont{Schatz}},
  \bibinfo{author}{\bibfnamefont{K.-L.} \bibnamefont{Kratz}}, \bibnamefont{and}
  \bibinfo{author}{\bibfnamefont{P.}~\bibnamefont{M\"{o}ller}},
  \bibinfo{journal}{PoS} \textbf{\bibinfo{volume}{NIC-IX}},
  \bibinfo{pages}{075} (\bibinfo{year}{2006}), \bibinfo{note}{and S. Gupta,
  private communication.}

\bibitem[{\citenamefont{{Timmes} and {Swesty}}(2000)}]{timmes.swesty:accuracy}
\bibinfo{author}{\bibfnamefont{F.~X.} \bibnamefont{{Timmes}}} \bibnamefont{and}
  \bibinfo{author}{\bibfnamefont{F.~D.} \bibnamefont{{Swesty}}},
  \bibinfo{journal}{Astrophys. J. Supp.} \textbf{\bibinfo{volume}{126}},
  \bibinfo{pages}{501} (\bibinfo{year}{2000}).

\end{thebibliography}

\end{document}